\documentclass[aip,pof,reprint,amsmath,amssymb]{revtex4-1}

\usepackage{graphicx}
\usepackage{dcolumn}
\usepackage{bm}
\usepackage{tikz}
\usepackage{subfig}
\usepackage{epstopdf}
\usepackage{tabularx}

\renewcommand{\vec}[1]{\mbox{\boldmath $ #1 $}}
\newcommand{\ensemble}[1]{\left\langle #1 \right\rangle}

\begin{document}

\preprint{AIP/123-QED}

\title[]{HPC realization of a controlled turbulent round jet using OpenFOAM\footnote{}}
\thanks{This paper was presented at \textbf{7th Open Source CFD International Conference} which took place in Hamburg, Germany on 24th \& 25th October 2013.}

\author{Asim Onder}
 \altaffiliation{}
\email{asim.onder@mech.kuleuven.be}
\author{Johan Meyers}%
\affiliation{
Department of Mechanical Engineering, Katholieke Universiteit Leuven
}%


\begin{abstract}
The present paper investigates high performance computing abilities of OpenFOAM for a low Reynolds number ($Re_D=2000$) axisymmetric jet subject to multiple zero net mass flux (ZNMF) actuators. First, parallel performance of OpenFOAM is tested by performing a scaling study up to $2048$ processors on a supercomputer of Flemish Supercomputer Center(VSC). Then, a method to improve the parallel efficiency is proposed. The method is based on developing a hybrid concept to calculate the statistical moments. This new concept combines ensemble and time averaging in order to allow data sampling in parallel. The motivation is obtaining a reduction in the walltime to collect turbulent statistics which is observed to be the dominating part in the ZNMF controlled jet flow. Employing this parallel statistical averaging approach in combination with regular grid partitioning parallelism, allowed us conducting DNS cases on $P=624$ processors with an overall speed-up of $S_e=540.56$ and a parallel efficiency of $E_e=0.87$. The parallelization using only grid partitioning exhibited inferior performance with $S_e=423.94$ and $E_e=0.68$.  In addition to develop a methodology to increase the parallel performance, we reduced the time step cost of the existing unsteady solver as well. To this end, an incremental projection method is implemented into OpenFOAM and a performance gain above $30\%$ is realized.
\end{abstract}

\maketitle

\section{\label{sec:Intro}Introduction}

Controlling turbulent round jet flows is of major interest because of technological relevance.  Typical applications are reducing the aircraft noise and improving the mixing efficiency in combustion and exhaust systems.  Zero net mass flux (ZNMF) actuation using synthetic jets is one of the popular methods used in this type of applications.\cite{tamburello08}  In the current work, we investigate ZNMF controlled jets by conducting direct numerical simulation (DNS) using the open source CFD toolbox OpenFOAM.\cite{Weller98}

In DNS all the relevant scales of turbulence need to be numerically well resolved. However, ZNMF actuators have a length scale that is an order magnitude lower than that of the main jet. Thus, resolving them on a computational grid designed primarily for the main jet is computational challenge.  Considering the second order spatial accuracy of finite volume discretization in OpenFOAM these conditions require an excessive number of grid elements to be employed. Hence, the scalability of the solver becomes very important.

In the present research effort, we performed a set DNS simulations of controlled round jets on a supercomputer of the Flemish Supercomputer Center (VSC).  The main focus of the study is on the development of methodologies that will allow us to realize these numerical experiments within a feasible time framework. The considered Reynolds number is low compared to the Reynolds numbers to be found in practical applications. Nevertheless, certain aspects of the low Reynolds number jet are also inherited by high Reynolds number jets.

In order to realize a large scale turbulent flow simulation on a supercomputer, first we use a method that improves the efficiency of parallelism by means of an optimal combination of ensemble averaging and grid partitioning based on the idea of Carati et al.\cite{Carati02} and further elaboration in Onder et al.\cite{Onder12}.  Using this approach, we are able to conduct DNS cases each using 624 processors.

Another parameter to be improved was the time step cost of the unsteady solver in OpenFOAM. To this end, we implemented an incremental projection scheme\cite{VanKan86} to replace the costly iterative algorithm in OpenFOAM's PisoFOAM. The coupled system of momentum and continuity equations are segregated with this non-iterative scheme without losing the second order accuracy of time integration.

The paper is organized as follows: In Sec. \ref{sec:comp} we introduced the flow problem and related computational details. Then, in Sec. \ref{sec:proj} the implemented projection algorithm is presented.
Furthermore, the parallel statistical averaging method to improve the parallel efficiency is discussed in Sec. \ref{sec:ensemble}. Some DNS results are presented in Sec. \ref{sec:DNS}. Finally, the conclusions are given in Sec. \ref{sec:conclusion}.

\section{\label{sec:comp} Numerical setup}
This section is devoted to the description of the details for conducted numerical experiments.  First, we will define the flow problem and then the computational details will follow.

\subsection{\label{sec:flowConfig} Flow configurations}

We consider a jet discharged into the free motionless ambient fluid through an axisymmetric orifice with a diameter $D$. The orifice is surrounded by solid walls. The jet at the orifice contains the full concentration of a scalar quantity which will be used in our mixing analysis. The governing equations are the incompressible Navier-Stokes equations

\begin{eqnarray}
\frac{\partial u_i}{\partial t}+\frac{\partial u_iu_j}{\partial x_i}&=&-\frac{\partial p}{\partial x_i}+\frac{1}{Re_D}\frac{\partial ^2 u_i}{\partial x_j \partial x_j},
\\
\frac{\partial u_i}{\partial x_i}&=&0
\label{eq:governingNS},
\end{eqnarray}
and a passive scalar transport equation:
\begin{eqnarray}
\frac{\partial c}{\partial t}+\frac{\partial u_i c}{\partial x_i}&=& \frac{1}{Re_D Sc}\frac{\partial ^2 c}{\partial x_j \partial x_j}
\label{eq:governingScalar},
\end{eqnarray}
where $u_i$ are the fluid velocity components, $c$ is the passive scalar, $Re_D$ denotes the Reynolds number based on the orifice diameter and jet orifice velocity $U_J$ and $Sc$ denotes the Schmidt number. All the cases considered in this work have a Reynolds number of $Re_D=2000$ and a Schmidt number of unity $Sc=1.0$. Numerical experiments are carried out using DNS approach where the intention is resolving all the scales of the turbulent flow.

\begin{figure*}[t]
%
%
\includegraphics[]{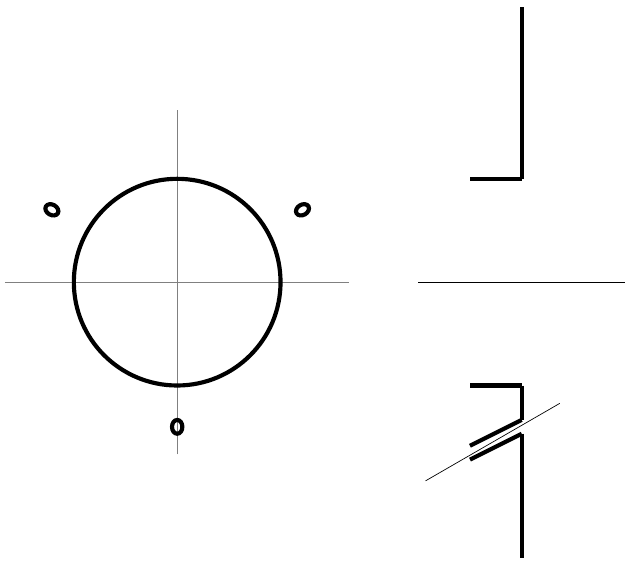}
\caption{\label{fig:act} Configuration of the actuators.}
\end{figure*}

We specified a uniform velocity $U_J$ in the core region and and laminar Blasius profile in the vicinity of the wall as used by a previous numerical work\cite{kim09}.  In order to control the main jet we employ three actuators distributed evenly in circumferential direction and placed $0.625D$ away from the jet centreline (cf. Figure \ref{fig:act} for an outline). The actuation is designed as a boundary condition on the wall and implemented as  harmonic oscillations
\begin{eqnarray}
u_{a}=U_a\sin(2.0\pi f_at)=U_a \sin(2.0\pi St_D U_J D^{-1}t).
\label{eq:control}
\end{eqnarray}
The actuation surface is an elliptic surface to resemble a real synthetic jet mounted in an inclined wall. This elliptic surface is the projection of the real experimental actuation plane making an angle of $\alpha=30^{\circ}$ with the jet centreline and it allows the same mass flux. Individual actuators have a momentum coefficient of  $C_{\mu}=0.0049$ and a unity actuation velocity to main jet velocity ratio $U_a/U_j=1$.  All the controls are in phase to manipulate the axisymmetric mode $m=0$ of the main jet.

The cases considered in this work are illustrated in Table~\ref{tab:cases}. We have designed two reference cases being subject to  relatively high perturbations on the main jet inlet flow that consist of white noise. To this end, we introduce temporally correlated noise with random spatial distribution as background disturbance on the inlet velocity while all the actuators are switched off. In the case $Base1$ we have observed a peak in the velocity spectra at a Strouhal number of $St_D=St_{pm}=0.33$. We assumed that this value is the preferred mode of the baseline jet and based on this value we selected four different control frequencies with $St_D=0.165$, $St_D=0.33$, $St_D=0.66$ and $St_D=1.32$. In the controlled cases the perturbation level on the baseline jet  is reduced to $u_J'=0.0015U_J$.

\begin{table*}
\caption{\label{tab:cases}Nomenclature and control parameters of investigated cases. ($St_D$):actuation frequency, ($u_J'/U_J$):background turbulence level, ($C_{\mu}$):momentum coefficient for individual actuators}

\begin{ruledtabular}
\begin{tabular}{lcccccc}
Cases                 & Base1          & Base2     & $Con0.5$    &     $Con1$     &   $Con2$     &    $Con4$ \\
\hline
$St_D$                & 0          & 0     &   0.165    &    0.33    &   0.66    &    1.32\\
$u_J'/U_J$            & 0.015     & 0.075  &   0.0015  &    0.0015  &   0.0015  &    0.0015\\
$C_{\mu}$      & 0  & 0 &  0.0049&  0.0049&  0.0049&  0.0049
\end{tabular}
\end{ruledtabular}

\end{table*}

\subsection{\label{sec:compDet} Computational details}

The open-source C++ library OpenFOAM v2.1.x \cite{Weller98} is employed in this study. A cylindrical computational domain is selected which extends $L=16D$ in axial direction and $R=8D$ in radial direction. A multiblock-structured grid using hexahedral elements with a square-shaped central block and O-grid building surrounding blocks is designed for calculations. The grid resolution is fixed to $130(x) \times 130(y) \times 475(z)$ elements for the core region and $470(r) \times 520(\theta) \times 475(z)$ elements for the O-grid region, which makes in total around  $125 \cdot 10^6$ elements. In order to represent accurately the synthetic jet--main jet interaction, a relatively higher circumferential resolution is used compared to the uncontrolled turbulent round jet simulations. Close to the jet exit,  the grid elements are clustered in the jet core region and in the shear layer. This cluster region expands radially moving downstream to resolve the jet spreading effectively. In the streamwise direction the refinements are applied towards the wall in order to resolve the oscillatory actuation dynamics.

Boundary conditions $\frac{\partial u_i}{\partial r}=0 $ and $\frac{\partial u_i}{\partial x}=0 $ correspond to lateral and outflow boundaries respectively. No inflow is allowed on the outflow boundary. In the framework of discretization the effect of actuation is projected on DNS grid using a Gaussian filter.

In OpenFOAM, the equations are discretized in space using a collocated finite volume method where all the variables are stored at the centroids of the grid elements. In the conducted simulations, the face values for the calculation of convective and diffusive terms are approximated using linear interpolation. The flux velocities on faces are modified using Rhie-Chow interpolation to prevent velocity-pressure decoupling. The nonlinear convective term is linearized by employing a second order extrapolation in time for the velocity flux. Temporal discretization and the projection method are discussed in the next chapter.

The time step size is fixed to $\delta t=0.004DU_J^{-1}$ which enables us to resolve one period of actuation in the highest frequency case, i.e. $Con4$, with 200 time steps. The initialization of simulations take $T_i=16DU_J^{-1}$. Afterwards we collect the statistics for a time interval of $T_s=320DU_J^{-1}$. For case $Con1$, the initial transient $T_i$ corresponds to 20 cycles of actuation and sampling time $T_s$ spans 400 cycles.

\section{\label{sec:proj} The incremental projection method}

The original projection methods by Chorin\cite{Chorin68,Chorin69} and Temam\cite{Temam68,Temam69} were developed as time marching techniques for evolutionary incompressible fluid problems. The name of the projection methods comes from the idea of projecting a
vector field onto a subspace of solenoidal vector fields.

The classical projection algorithms may be classified according to the pressure extrapolation they employ in the subproblems. According to this classification the two main classes are: the non-incremental projection method, and the incremental projection method. The former uses simply no pressure gradient (zero-order approximation) in the momentum equation (sometimes called Burger’s) step. In contrast, the latter method employs a first order pressure extrapolation by using the pressure from the previous time step as the approximated pressure and corrects it in the final step by incrementing with the pressure-correction term obtained in the projection step. This class has been found more attractive due to increase in the accuracy without any extra computational demand compared to the non-incremental one. The incremental method was first used by Goda\cite{Goda79}. Then, Van Kan\cite{VanKan86} proposed a second-order accurate scheme which combines the incremental algorithm with semi-implicit Crank-Nicolson time discretization.

Van Kan's scheme is the method of choice in this work to replace the costly iterative PISO algorithm in OpenFOAM. In order to ease the description we will present the algorithm in spatially continuous framework. Using a Crank-Nicolson scheme with a semi-implicit convective term for the time integration, the basic
steps of the incremental projection scheme read as follows:

 \begin{enumerate}
  \item Momentum step: Given $u_i^n$ and $p^n$ from previous time step, solve for $\tilde{u}_i^{n+1}$ from
  \begin{equation}
\frac{\tilde{u_i}^{n+1} - u_i^n }{\delta t} +
\left(\frac{3}{2} u_j^{n} - \frac{1}{2} u_j^{n-1} \right)\frac{\partial \tilde{u_i}^{n+1/2}}{\partial x_j}- \frac{1}{Re_D}\frac{\partial ^2 \tilde{u_i}^{n+1/2}}{\partial x_j \partial x_j} = -\frac{\partial p^{n}}{\partial x_i}
\label{eq:proj1}
\end{equation}
where
\begin{equation*}
\tilde{u_i}^{n+1/2}=\frac{1}{2}\left(\tilde{u_i}^{n+1}+u_i^{n}\right).
\label{eq:proj1}
\end{equation*}
  \item Projection step: Perform the projection
by solving first
\begin{equation}
\frac{\partial ^2 \left(p^{n+1}- p^{n}\right)}{\partial x_i \partial x_i}=\frac{1}{\delta t}\frac{\partial \tilde{u_i}^{n+1}}{\partial x_i}
\end{equation}
then, updating the velocity
\begin{equation}
u_i^{n+1}= \tilde{u_i}^{n+1}-\delta{t}\frac{\partial \left(p^{n+1}- p^{n}\right)}{\partial x_i}.
\end{equation}
\end{enumerate}

 The coupled system of momentum and continuity equations are segregated with this scheme. The segregation yields an additional second order error in time. In contrast to OpenFOAM's $pisoFoam$ solver which has only a first order approximation to the nonlinear convection velocity (Rhie-Chow corrected convective flux in discretized form) we empoyed a second order approximation in time by using two previous time steps. Combining this linearization method with the Crank-Nicolson time discretization yields a second order temporal accuracy overall. Hence, the order of accuracy of the segregation is consistent with the time marching scheme.

 Following the projection scheme, each time step a linear system of convection--diffusion like equations for each velocity component have to be solved. This is done by using the biconjugate gradient iterative solver with a  diagonal incomplete LU preconditioner up to a solution tolerance of $10^{-9}$. The second and more expensive step of the projection scheme requires solving a discretized Poisson equation for the pressure. As the computational grid contains some regions with slightly non-orthogonal elements we employ two extra iterations to reduce the non--orthogonality effect.  In total, we solve the Poisson iteration three times, using a conjugate gradient solver with a geometric-algebraic multigrid preconditioner up to a tolerance of $10^{-10}$ in the final step.

We benchmarked the new projection solver with the existing Piso solver in OpenFOAM. For these tests, we employed 128 processors using grid partitioning paralelism with scotch method on the $Tier1$ supercomputer of Flemish Supercomputer Center (VSC). $Tier1$ of VSC has a total number of 8448 computer cores equipped with Intel Sandy Bridge microprocessor technology and FDR Infiniband Mellanox communication network. For the benchmark case, we specified two iterations for Piso solver. Both solvers also included the solution of the passive scalar equation at the end of each time step as well. As a result, projection solver delivered a time step cost of $49.1$ s and piso solver covered a time step in $71.5$ s. According to these results, the performance gain with the projection solver was over $30\%$.

\section{\label{sec:ensemble} Parallel statistical averaging}
In this section we discuss the method we employed to improve the scalability of OpenFOAM in turbulent flow simulations. In direct numerical simulations of turbulent flows, the chaotic and strongly fluctuating velocity fields (in space and time) associated to turbulence, are directly represented in the simulations. For practical purposes, these three-dimensional time-varying velocity fields need to be averaged, yielding mean-velocity profiles, Reynolds stresses, and similar turbulent-flow quantities. Formally, averaging in turbulence is defined using the average over an ensemble of statistical independent realizations of the same flow. For turbulent flow systems that are statistically stationary, this ensemble average is in practice replaced by averaging the solution in time. To this end, simulations are first run for some period of time $T_i$ in which the flow starts from an initial condition (usually constructed for some part using random velocities), and subsequently settles into a statistical equilibrium. Afterwards, averaging in time is performed over a period of time $T_s$ which needs to be sufficiently long to acquire statistically converged mean-flow quantities.

In order to be able to parallelize the averaging of the turbulent solution in LES or DNS of statistically stationary flows, we propose to partially resort back to the definition of an ensemble average, i.e. we propose to build a set of $R$ statistical independent realizations of the same flow, that can be simulated independently (in parallel), with an interval for time averaging that is now reduced to $T_s/R$. This averaging operation reads as follows:
\begin{equation}
\ensemble{\vec{u}(\vec{x})} \approx \frac{1}{R} \sum_{r=1}^R  \frac{R}{T_S} \int_{t_0}^{t_0+T_S/R} \vec{u}^{\{r\}}(\vec{x},t) \ {\rm d}t, \label{eq:ensembleRT}
\end{equation}
This idea is fairly simple, and was, e.g. already proposed in a slightly different context by Carati et al\cite{Carati02}. The major drawback for the current case is that all $R$ simulations need to be initialized, and require a start-up run during which the initial velocity field evolves into statistical equilibrium. The required time $T_i$ is typically of the order of a few through-flow times of the simulation domain; a more precise discussion is provided in Onder et al.\cite{Onder12}. Taking this non-parallelizible initialization time $T_i$ into account in the proposed parallelization of ensemble averaging, we end up with a parallel speed-up, and parallel efficiency respectively of
\begin{eqnarray}
S_e(p,R) &=& \frac{\mathbf{T}^w_1}{\mathbf{T}^w_{P}}=  \frac{T_i+ T_s}{T_i + T_s/R} S_g(p) = R \frac{k+1}{R k+1} S_g(p) \\
\label{eq:s1}
E_e(p,R) &=&  \frac{\mathbf{T}^c_1}{\mathbf{T}^c_P}=  \frac{S_e(p,R)}{pR},
\label{eq:e1}
\end{eqnarray}
with $k=T_i/T_s$, $S_g(p)$ is the speed-up due to grid partitioning and where we introduce $\mathbf{T}^w$, and $\mathbf{T}^c$ respectively as the \emph{wall-time}, and the \emph{cpu-time} of computations. It is appreciated that the efficiency of the proposed parallelization becomes very poor for $R k \gtrsim 1$. Moreover, unless $k \ll 1$, the potential for speed-up is rather limited. Given $T_i=16DU_J^{-1}$ and  $T_s=320DU_J^{-1}$ (cf. Sec. \ref{sec:comp}), we have $k=0.05$ for our DNS cases. This low value of $k$ allowed us to efficiently  employ parallel statistical averaging in our studies.

\begin{table*}
\caption{\label{tab:scaling} Parallel performance of OpenFOAM for $Con1$ case. Linear speed-up is assumed for $p=64$ case.}
\begin{ruledtabular}
\begin{tabular}{lcccccccc}
p & $64$& $128$ & $156$&$256$&$512$&$624$&$1024$&$2048$ \\
\hline
time step cost$(s)$ & 98.7 & 49.1 & 40.9 & 27.7 & 16.3&14.7 & 11.9& 12.1 \\
$S_g(p)$ & 64 & 128.65 & 154.44 & 228.04 & 387.63 & 423.94&530.82 & 522.04 \\
$E_g(p)=S_g(p)/p$ & 1 & 1.005 & 0.99 & 0.89 & 0.76 &0.68& 0.52 & 0.25
\end{tabular}
\end{ruledtabular}
\end{table*}

\begin{figure*}[t]
\includegraphics[]{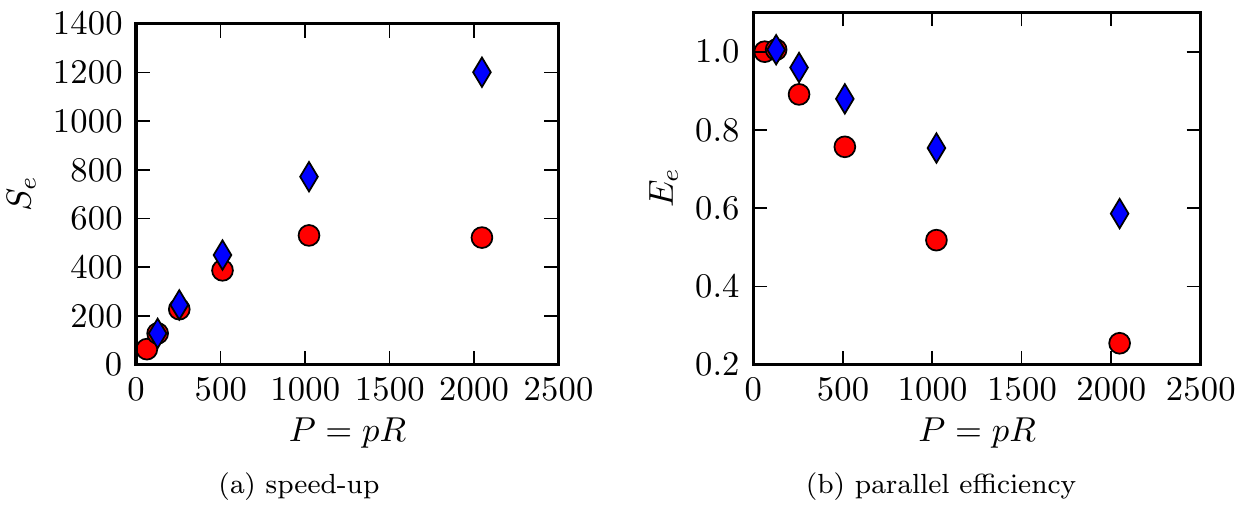}
\caption{\label{fig:speed-up} Parallel performance for two cases. ($\circ$) :  $S_e(p,1)=S_g(p)$ (only grid partitioning) ; ($\diamond$) :  $S_e(p=128,R=1-16)$.}
\end{figure*}

In order to find the right balance between $p$ and $R$ we conducted a scaling study covering $1000$ time steps of case $Con1$. The results are reported in Table \ref{tab:scaling}. Using these results, we first explored the potential of parallel statistical averaging by investigating an experimental case with $p=128$ and $R=1-16$. The comparison of the speed-up and efficiency of this case with the ones of regular grid partitioning is illustrated in Fig. \ref{fig:speed-up}. We can clearly see in these figures that the addition of ensemble averaging concept on top of grid partitioning parallelism improves the parallel performance.

According to the data, we see that up to $p=156$ processors the code exhibit linear speed-up. Therefore, for this study we selected a configuration of $p=156$ and $R=4$ for our DNS studies. This configuration delivers an overall speed-up of $S_e=540.56$ and $E_e=0.87$ on $P=624$ processors, where $S_g(624)=423.94$ and $E_g(624)=0.68$ only. The improved parallelism with parallel statistical averaging is clear. Moreover, we have to note that a potential for further improvement exists if number of branches $R$ is increased. As mentioned before, we designed six different test cases (cf. Table \ref{tab:cases}) and ran them simultaneously making a total number of $3744$ processors. Hence, we limited $R$ to $4$ in each case and didn't exploit the full potential of parallel statistical averaging.

\section{\label{sec:DNS} DNS results}
\begin{figure*}[t]
\includegraphics[]{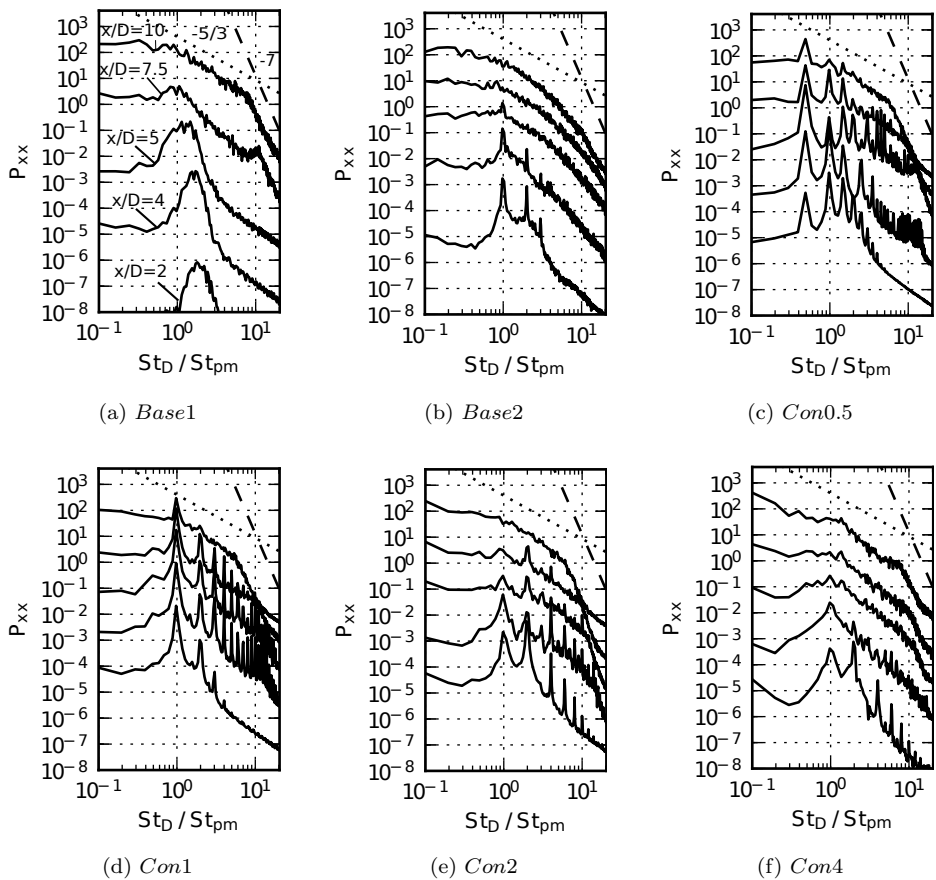}
\caption{\label{fig:FFTs} Power spectra of centreline velocity signals with normalized frequency axis using the preferred mode value $St_{pm}=0.33$. Each spectrum is shifted one decade starting from the signals on $x/D=10$ }
\end{figure*}
\begin{center}
\begin{figure*}
\includegraphics[]{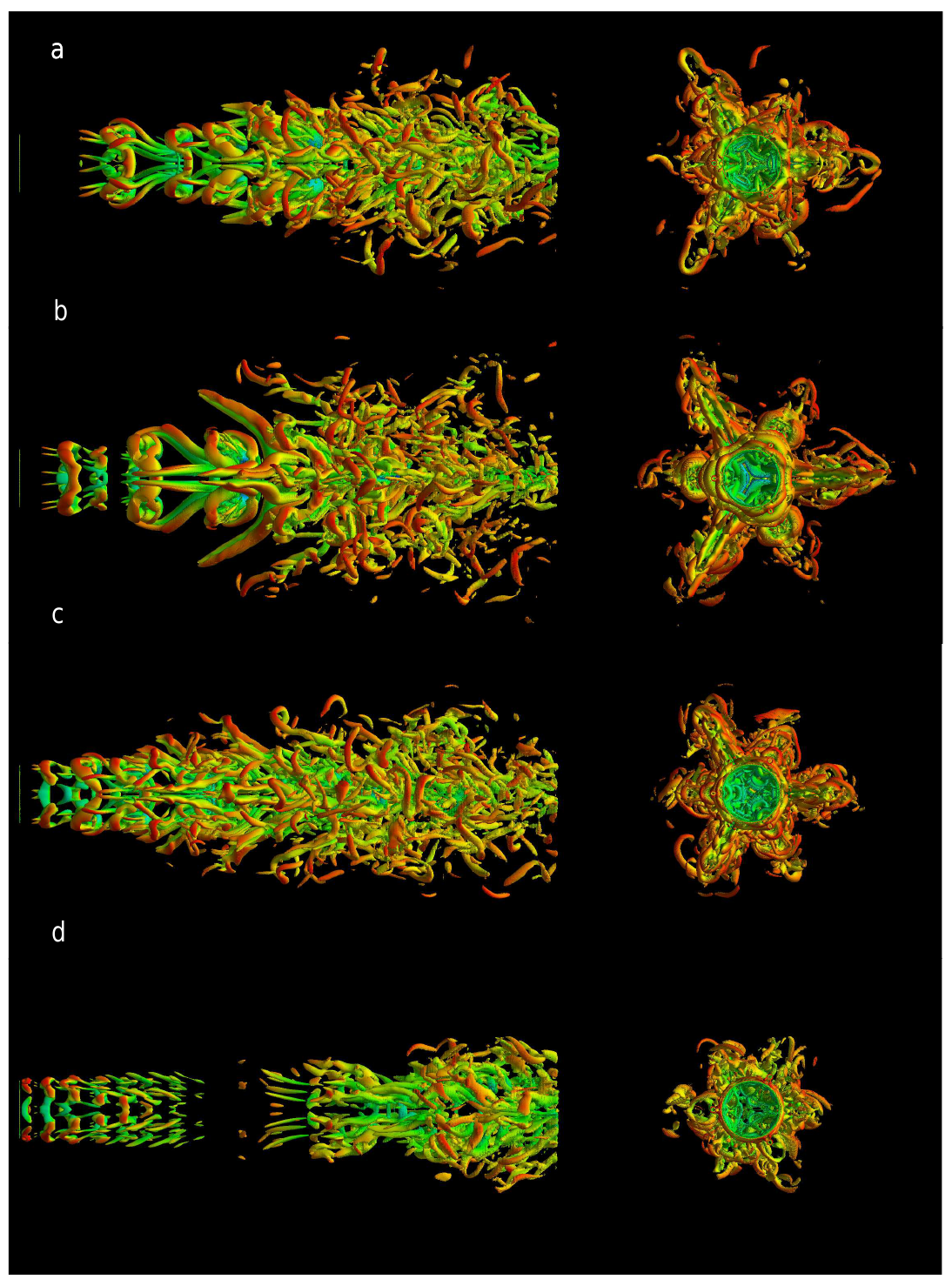}
\caption{\label{fig:Qall}Visualization of vortical structures by isosurfaces of Q criterion. Coloring by velocity magnitude for cases (a) $Con0.5$; (b) $Con1$; (c) $Con2$; (d) $Con4$; (left) streamwise view in non-actuation plane; (right) transverse view in $x=0$}
\end{figure*}
\end{center}
In this section we present some results from DNS experiments. First, we have calculated the power spectra of streamwise component of fluctuating velocity signals at particular locations on the jet axis ($r/D=0$). The results are reported in Fig. \ref{fig:FFTs}. The dotted lines in these figures correspond to $-5/3$ slope of the Kolmogorov spectrum in the inertial region and the dashed lines correspond to $-7$ slope of the Pao spectrum characterizing the dissipation range\cite{pope2000}. The difference in the controlled cases compared to the unactuated cases is the display of distinctive peaks corresponding to actuation frequencies and their subharmonics. These peaks represent the enhancement of coherent motion under the effect of the actuation. For cases $Con0.5$ and $Con1$ spectral peaks due to actuation diminish very slowly when moving downstream. Even at $x/D=10$ they are evident. In contrast, spectra of $Con2$ and $Con4$ at $x/D=10$ is very similar to the ones observed in baseline cases and demonstrate no distinctive peaks on wide-scale turbulence spectra.  Moreover, another obvious distinction in these high frequency control cases is the additional peaks at half values of the actuation frequencies which are the result of vortex pairings events.

We illustrated next the resulting vortical structures in Fig.~\ref{fig:Qall} using $Q$ criterion.\cite{Hunt88}. The evolution of the vortical structures strongly depends on the control frequency. Close to the orifice the shed frequency of the Kelvin-Helmholtz vortex rings are locked to the actuation frequency. As in the $Con1$ case the actuation is in a frequency regime which is the closest to the natural jet instability frequencies\cite{crow70}, the most enhanced coherence and large scale structure development take place. For actuation with the highest frequency considered in this work, i.e. $Con4$, there is a lot less coherent structure growth compared to the other cases.

In Fig.~\ref{fig:FluxQ} the results for entrainment of the ambient fluid is plotted. Entrainment is a commonly used parameter in mixing studies as the first stage of mixing is entraining the surrounding ambient fluid into the turbulent jet core. We observe that the entrainment is enhanced significantly by actuated cases $Con0.5$, $Con1$ and $Con2$. $Con1$ case delivers again superior results compared to other controlled cases.

The decay of passive scalar concentration on the jet centreline is a more direct indicator of mixing efficiency. In Fig.~\ref{fig:C_c} the development of the reciprocal of centreline scalar concentration $C_c=\left\langle c \right\rangle_{r=0}$ is shown for all cases. Here we see directly that actuation in preferred mode is very effective in terms of mixing efficiency.

\begin{figure*}[t]
\includegraphics[]{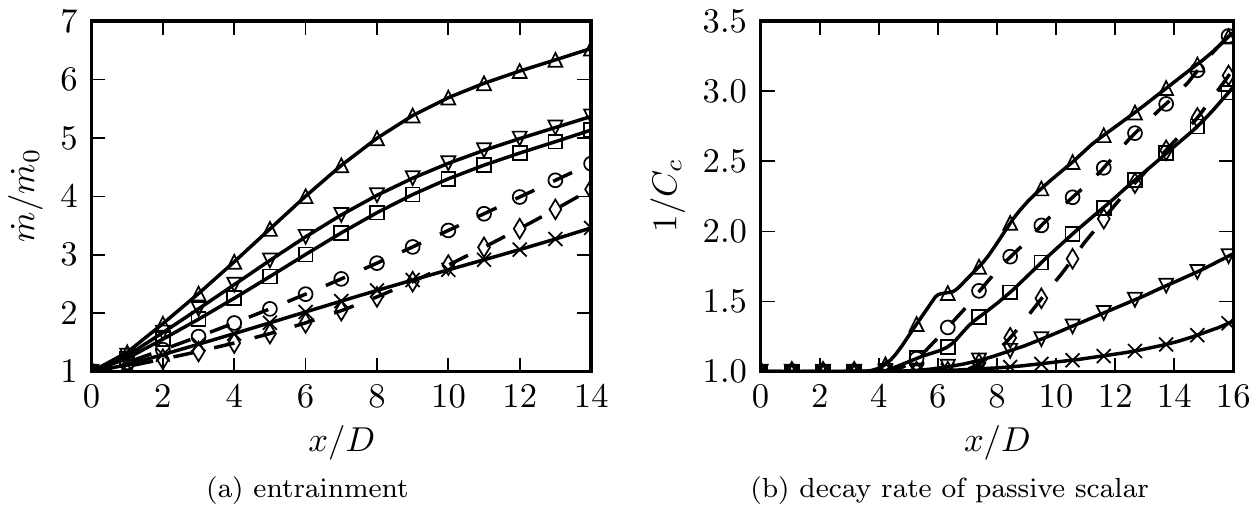}
\caption{\label{fig:} The variation of entrainment and passive scalar decay rate with axial distance. (- -$\diamond$): $Base1$; (- -$\circ$): $Base2$ ; (---$\square$): $Con0.5$;  (---$\triangle$): $Con1$; (---$\nabla$): $Con2$; (---$\times$): $Con4$.}
\end{figure*}

\section{\label{sec:conclusion}Concluding remarks}
In this work we tested and improved the high performance computing capabilities of OpenFOAM for a low Reynolds number ($Re_D=2000$) axisymmetric jet subject to multiple ZNMF actuators. Our motivation was first increasing the limited scalability of OpenFOAM by using parallel statistical averaging in combination with grid partitioning parallelism. This approach allowed us conducting DNS cases on $P=624$ processors with an overall speed-up of $S_e=540.56$ and and parallel efficiency of $E_e=0.87$. It is shown that the parallelization using only grid partitioning delivered inferior performance with $S_e=423.94$ and $E_e=0.68$ for $P=624$.  We also showed that in this configuration a potential for further improvement exists if the number of members in the ensemble is increased.

Additionally, we aimed to reduce the time step cost of the existing unsteady solver. Hence, an incremental projection method is implemented and a performance gain above $30\%$ is achieved.

\bibliographystyle{abbrv}

\begin{thebibliography}{10}

\bibitem{Carati02}
D.~Carati, M.~Rogers, and A.~Wray.
\newblock {Statistical ensemble of large-eddy simulations}.
\newblock {\em {Journal of Fluid Mechanics}}, {455}:{195--212}, {MAR 25}
  {2002}.

\bibitem{Chorin68}
A.~J. Chorin.
\newblock Numerical solution of the {N}avier-{S}tokes equations.
\newblock {\em Mathematics of computation}, 22(104):745--762, 1968.

\bibitem{Chorin69}
A.~J. Chorin.
\newblock On the convergence of discrete approximations to the
  {N}avier-{S}tokes equations.
\newblock {\em Mathematics of Computation}, 23(106):341--353, 1969.

\bibitem{crow70}
S.~C. Crow and F.~H. Champagne.
\newblock Orderly structure in jet turbulence.
\newblock {\em Journal of Fluid Mechanics}, 48:547--591, 7 1971.

\bibitem{Goda79}
K.~Goda.
\newblock A multistep technique with implicit difference schemes for
  calculating two-or three-dimensional cavity flows.
\newblock {\em Journal of Computational Physics}, 30(1):76--95, 1979.

\bibitem{Hunt88}
J.~C. Hunt, A.~Wray, and P.~Moin.
\newblock Eddies, streams, and convergence zones in turbulent flows.
\newblock In {\em Studying Turbulence Using Numerical Simulation Databases, 2},
  volume~1, pages 193--208, 1988.

\bibitem{kim09}
J.~Kim and H.~Choi.
\newblock Large eddy simulation of a circular jet: effect of inflow conditions
  on the near field.
\newblock {\em Journal of Fluid Mechanics}, 620:383--411, 1 2009.

\bibitem{Onder12}
A.~Onder, P.~Wu, and J.~Meyers.
\newblock Improving speed-up and efficiency in simulation of stationary
  turbulent flows by parallelization of statistical averaging.
\newblock In {\em Proceedings of the 9th International ERCOFTAC Symposium on
  Engineering Turbulence Modeling and Measurements (ETMM9)}, June 2012.

\bibitem{pope2000}
S.~Pope.
\newblock {\em Turbulent flows}.
\newblock Cambridge University Press, 2000.

\bibitem{tamburello08}
D.~A. Tamburello and M.~Amitay.
\newblock Active control of a free jet using a synthetic jet.
\newblock {\em International Journal of Heat and Fluid Flow}, 29(4):967 -- 984,
  2008.

\bibitem{Temam68}
R.~Temam.
\newblock Une m{\'e}thode d'approximation de la solution des {\'e}quations de
  {N}avier-{S}tokes.
\newblock {\em Bulletin de la Soci{\'e}t{\'e} Math{\'e}matique de France},
  96:115--152, 1968.

\bibitem{Temam69}
R.~Temam.
\newblock On the approximation of the solution of {\ 'e} {N}avier-{S}tokes
  equations by m {\' e} method of fractional steps (i).
\newblock {\em Archive for Rational Mechanics and Analysis}, 32(2):135--153,
  1969.

\bibitem{VanKan86}
J.~van Kan.
\newblock A second-order accurate pressure correction scheme for viscous
  incompressible flow.
\newblock {\em SIAM J. Sci. Stat. Comput.}, 7(3):870--891, July 1986.

\bibitem{Weller98}
H.~G. Weller, G.~Tabor, H.~Jasak, and C.~Fureby.
\newblock A tensorial approach to computational continuum mechanics using
  object-oriented techniques.
\newblock {\em Comput. Phys.}, 12:620--631, November 1998.

\end{thebibliography}

\end{document}